\documentclass{WileyMSP-template}
\usepackage[skip=15pt plus1pt, indent=0pt]{parskip}
\usepackage{lineno}
\usepackage[normalem]{ulem}
\usepackage{xcolor}
\usepackage{upgreek}

\begin{document}
\pagestyle{fancy}
\rhead{\includegraphics[width=2.5cm]{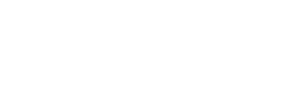}}

\title{Spin-orbit torques and spin Hall magnetoresistance generated by twin-free and amorphous Bi\textsubscript{0.9}Sb\textsubscript{0.1} topological insulator films  
}
\maketitle

\author{Federico Binda,*}
\author{Stefano Fedel,}
\author{Santos Francisco Alvarado,}
\author{Paul No{\"e}l,}
\author{Pietro Gambardella*}

\begin{affiliations}
Dr. F. Binda, S. Fedel, Dr. S. F. Alvarado, Dr. P. No{\"e}l, Prof. P. Gambardella\\
Address: Department of Materials, ETH Zurich, CH-8093 Zurich, Switzerland\\ 
Email Address: federico.binda@mat.ethz.ch; pietro.gambardella@mat.ethz.ch
\end{affiliations}


\begin{abstract}

Topological insulators have attracted great interest as generators of spin-orbit torques (SOTs) in spintronic devices. Bi\textsubscript{1-x}Sb\textsubscript{x} is a prominent topological insulator that has a high charge-to-spin conversion efficiency. However, the origin and magnitude of the SOTs induced by current-injection in Bi\textsubscript{1-x}Sb\textsubscript{x} remain controversial. Here we report the investigation of the SOTs and spin Hall magnetoresistance resulting from charge-to-spin conversion in twin-free epitaxial layers of Bi\textsubscript{0.9}Sb\textsubscript{0.1}(0001) coupled to FeCo, and compare it with that of amorphous Bi\textsubscript{0.9}Sb\textsubscript{0.1}. We find a large charge-to-spin conversion efficiency of 1 in the first case and less than 0.1 in the second, confirming crystalline Bi\textsubscript{0.9}Sb\textsubscript{0.1} as a strong spin injector material. The SOTs and spin Hall magnetoresistance are independent of the direction of the electric current, indicating that charge-to-spin conversion in single-crystal Bi\textsubscript{0.9}Sb\textsubscript{0.1}(0001) is isotropic despite the strong anisotropy of the topological surface states. Further, we find that the damping-like SOT has a non-monotonic temperature dependence with a minimum at 20~K. By correlating the SOT with resistivity and weak antilocalization measurements, we conclude that charge-spin conversion occurs via thermally-excited holes from the bulk states above 20~K, and conduction through the isotropic surface states with increasing spin polarization due to decreasing electron-electron scattering below 20~K.

\end{abstract}

\clearpage

{\sloppy

\section{Introduction}
Spin-orbit torques (SOTs) are current-induced magnetic torques that permit the electrical manipulation of the magnetization in thin film heterostructures \cite{G117} and have application in low-power magnetic memory and logic devices \cite{G231, luo2020current, krizakova2022spin}. 
\color{black} SOTs were initially discovered in heavy metal/ferromagnet bilayers
 \cite{ G117, ando2008electric, G118, miron2011perpendicular, liu2012spin, G139, G19, G50, kim2013layer} 
and noncentrosymmetric magnetic semiconductors \cite{chernyshov2009evidence, kurebayashi2014antidamping}.
In these materials, the most prominent charge-to-spin conversion mechanisms are the spin Hall effect and the Rashba-Edelstein effect, usually associated with bulk and interfacial spin-orbit coupling, respectively \cite{G117}. 
\color{black} Since the first SOT studies, the quest for larger charge-to-spin conversion efficiency has led to exploring new materials as SOT generators. Topological insulators (TIs) are among the most promising as they possess a bulk band gap and conducting surface states (SSs) in which the direction of the electrons' wavevector and spin are coupled to each other by the spin-orbit interaction \cite{G4, G136}. This so-called spin-momentum locking allows for a fully spin-polarized current to flow at the interface of the TI.
This feature has stimulated much interest in exploring TIs as spin injector materials for generating SOTs in ferromagnetic heterostructures, leading to several reports of a high charge-to-spin conversion efficiency 
\cite{G128, G126, G197, G206, G205, G211, G220, G219, G222, binda2021spin}. 

Most of the TIs studied in the context of SOTs are bismuth chalcogenide compounds, such as Bi\textsubscript{2}Se\textsubscript{3}, Bi\textsubscript{2}Te\textsubscript{3}, and (Bi\textsubscript{1-x}Sb\textsubscript{x})\textsubscript{2}Te\textsubscript{3}.
The chalcogen elements Se and Te, however, are highly volatile and reactive, causing interdiffusion in heterostructures \cite{greenwood2012chemistry, house2015descriptive, ambrose1968vapor}.
Their growth as thin films further requires an overpressure of Se and Te, with a consequent limitation in stoichiometry accuracy \cite{podpirka2020role}. As a consequence, the bismuth chalcogenides often present metallic bulk conductivity due to the presence of Se or Te vacancies \cite{G221, jash2021imaging}. The Bi\textsubscript{1-x}Sb\textsubscript{x} compound with 0.07$<$x$<$0.22 is a TI that avoids many of these issues \cite{G4, G52, G201}. In particular, Bi\textsubscript{0.9}Sb\textsubscript{0.1} possesses a narrow band gap and carries topological SSs with a six-fold rotational symmetry in the (0001) plane, as shown by angle-resolved photoemission \cite{G201, G283, G265, G275}. 
\color{black}
Several works investigated the SOTs in Bi\textsubscript{1-x}Sb\textsubscript{x} heterostructures with $\rm{x}\approx 0.1$ \cite{G126, G197, G256, G259, G254,  G270, G261, khang2022nanosecond},
including current-induced magnetization switching \cite{G126, G256, G259, G254, G270, khang2022nanosecond}. However, the reported charge-to-spin conversion efficiency varies from close to 0 \cite{G261, sharma2021light} to 52 \cite{G126} depending on the deposition method, crystalline and interface quality of the films, and techniques used to measure the SOTs. The magnitude of the SOTs in Bi\textsubscript{1-x}Sb\textsubscript{x} heterostructures and their relationship to crystalline order thus remain highly debated.

The origin of charge-to-spin conversion in Bi\textsubscript{0.9}Sb\textsubscript{0.1} is also controversial. 
\color{black}
The large SOTs of epitaxial Bi\textsubscript{0.9}Sb\textsubscript{0.1} have been associated with the spin-momentum locked SSs \cite{G126}. However, the small band gap of Bi\textsubscript{0.9}Sb\textsubscript{0.1} allows for thermally excited carriers to participate in the electric conduction, whereas charge doping by defects can shift the Fermi level away from the gap. Hence, the spin-Hall and Rashba Edelstein effects due to bulk carriers can also contribute to charge-spin conversion and generation of SOTs \cite{G197, gao2019semi, G284}. The actual role of the SSs in generating the SOTs has been questioned in recent studies focusing on polycrystalline Bi\textsubscript{1-x}Sb\textsubscript{x}  films \cite{G197, gao2019semi}. Such films, deposited by magnetron sputtering \cite{ G197, G256, G259, G254, G270, G261, G255}, are very promising for applications but suffer from significant variations in film quality and imperfect crystal structure \cite{panjan2020review}. Moreover, crystal twinning often appears in rhombohedral epitaxial films grown along the unit cell diagonal, as in the case of Bi\textsubscript{0.9}Sb\textsubscript{0.1}(0001) \cite{ G276, G251, G257,  G255, G253,  G271}. The presence of defects, crystal grains, and twinning makes it difficult to compare the SOT generated by Bi\textsubscript{0.9}Sb\textsubscript{0.1} in different studies and also to pin down the origin of charge-to-spin conversion, e.g., by correlating the magnetotransport properties with the band structure of crystalline Bi\textsubscript{0.9}Sb\textsubscript{0.1} \cite{G201, G265, G275, baringthon2022topological, G262}.

Here, we investigate the generation of SOT and spin Hall magnetoresistance in twin-free epitaxial and amorphous Bi\textsubscript{0.9}Sb\textsubscript{0.1}/FeCo bilayers in order to shed light on the magnitude and origin of charge-to-spin conversion in single-crystal and disordered Bi\textsubscript{0.9}Sb\textsubscript{0.1}. 
\color{black}
This study includes SOT and magnetoresistance measurements as a function of crystal orientation and temperature in order to distinguish the bulk and SS contributions to charge-spin conversion. 
\color{black}

We developed a simple one-step method to grow twin-free Bi\textsubscript{0.9}Sb\textsubscript{0.1}(0001) thin films of high crystalline quality using molecular beam epitaxy (MBE) on BaF\textsubscript{2}(111) substrates. Our single layer epitaxial Bi\textsubscript{0.9}Sb\textsubscript{0.1} films present the same transport properties as crystalline Bi\textsubscript{1-x}Sb\textsubscript{x}, in particular the increasing resistance with temperature and the concomitant electron and hole conduction \cite{G269, G287}. The charge-to-spin conversion efficiency estimated for the damping-like and field-like SOT in the epitaxial sample are close to 1 and 0.2 at room temperature, and are reduced by 90\% and 50\% in the amorphous sample, respectively. The corresponding spin Hall magnetoresistance is 0.25\% and 0.07\%. 
\color{black}
These results show that the charge-to-spin conversion efficiency of crystalline Bi\textsubscript{0.9}Sb\textsubscript{0.1} is large compared to heavy metals, but significantly smaller than reported for epitaxial MnGa/Bi\textsubscript{0.9}Sb\textsubscript{0.1}($\bar{1}$2$\bar{1}$6) bilayers grown on GaAs(001) \cite{G126}. We comment on this discrepancy later in the manuscript. The comparison between twin-free single-crystal Bi\textsubscript{0.9}Sb\textsubscript{0.1} and  amorphous Bi\textsubscript{0.9}Sb\textsubscript{0.1} shows that structural disorder has a strong negative impact on the charge-to-spin conversion efficiency.

We further analyzed the SOT efficiency as a function of temperature in twin-free epitaxial Bi\textsubscript{0.9}Sb\textsubscript{0.1}. We find that the effective spin Hall conductivity decreases from 2.2 to $1.4\times 10^5$~$(\Omega \rm{m})^{-1}$ from 300 to 20~K, indicating that thermally-excited bulk carriers are the main source of SOTs. Moreover, the SOT efficiency is isotropic for current injection along different crystallographic directions at any probed temperature, despite the anisotropic surface and bulk band structure. These observations indicate that the carriers responsible for the SOTs originate from the isotropic T point of the Brillouin zone. Additionally, below 20~K we observe prominent weak antilocalization effects and an upturn of the SOT efficiency, which we associate with increased spin polarization and conduction by the isotropic SSs around the $\bar{\Gamma}$ point relative to the bulk. Our results provide a consistent description of charge-spin conversion phenomena in epitaxial and disordered Bi\textsubscript{0.9}Sb\textsubscript{0.1} thin films, as required for implementing this material in spintronic devices. 
\color{black}

\clearpage

\begin{figure}
  \includegraphics[width=\linewidth]{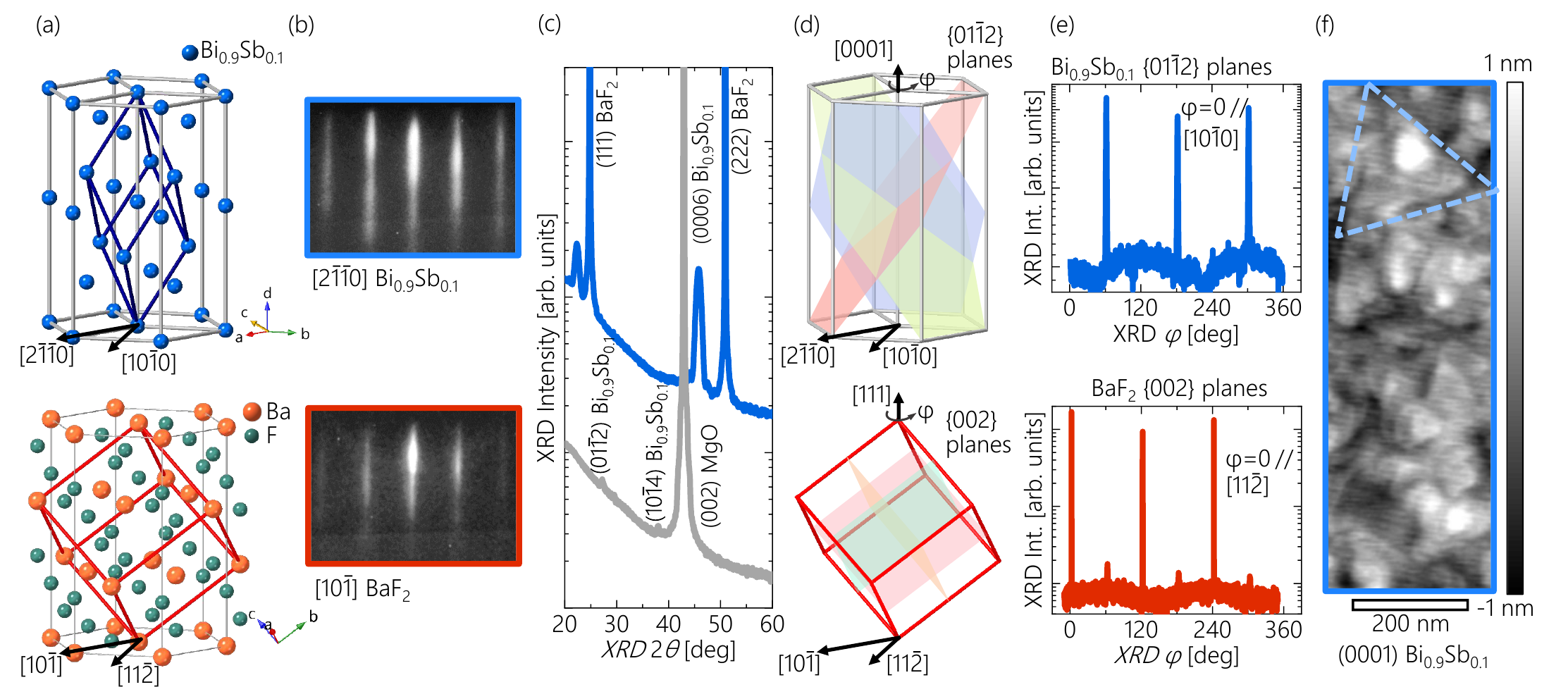}
  \caption{
  Epitaxial relationship and structural characterization of Bi\textsubscript{0.9}Sb\textsubscript{0.1} grown on BaF\textsubscript{2}(111). (a) Unit cell of rhombohedral Bi\textsubscript{0.9}Sb\textsubscript{0.1} (top) and  cubic BaF\textsubscript{2} (bottom). 
  The orientation of the  Bi\textsubscript{0.9}Sb\textsubscript{0.1} unit cell corresponds to one of the two possible twinning domains, and its conventional cell, in dark gray, is described in the hexagonal lattice system with four axis-notation. (b) RHEED patterns of the Bi\textsubscript{0.9}Sb\textsubscript{0.1} film and of the BaF\textsubscript{2}(111) substrate prior to deposition. (c) XRD 2$\theta$/$\omega$ scans of the twin-free and amorphous Bi\textsubscript{0.9}Sb\textsubscript{0.1} bilayer samples. (d) Unit cells showing the crystallographic planes detected in the XRD $\varphi$ scans. (e) XRD azimuthal $\varphi$ scans showing the three-fold rotational symmetry of Bi\textsubscript{0.9}Sb\textsubscript{0.1} (top) and BaF\textsubscript{2} (bottom), which proves the existence of a single twinning domain. (f) AFM image of an uncapped twin-free Bi\textsubscript{0.9}Sb\textsubscript{0.1} film. The dashed contour highlights the triangular-shaped features of the film, all similarly oriented, consistently with the growth of a single crystal domain.
  }
  \label{fig:boat1}
\end{figure}

\clearpage

\section{Results and Discussion}
\subsection{Sample growth and structural characterization}

We fabricated two different ferromagnetic heterostructures grown by MBE (see Experimental Section): {(i)~BaF\textsubscript{2}(111)/Bi\textsubscript{0.9}Sb\textsubscript{0.1}[100~\AA]/FeCo[17~\AA]/AlO\textsubscript{x}} and  {(ii)~MgO(001)/FeCo[17~\AA]/Bi\textsubscript{0.9}Sb\textsubscript{0.1}[100~\AA]/AlO\textsubscript{x}}. 
The first consists of the twin-free epitaxial Bi\textsubscript{0.9}Sb\textsubscript{0.1} film, grown with the (0001) orientation on the BaF\textsubscript{2}(111) substrate. 
The second consists of its amorphous counterpart deposited on the MgO(001) substrate with inverted layer stacking to inhibit epitaxial growth.
We refer to these samples as twin-free Bi\textsubscript{0.9}Sb\textsubscript{0.1} bilayer and amorphous Bi\textsubscript{0.9}Sb\textsubscript{0.1} bilayer, respectively.
Additionally, as control samples, we deposited single films of twin-free epitaxial Bi\textsubscript{0.9}Sb\textsubscript{0.1}[100~\AA] on BaF\textsubscript{2}(111), referred to as twin-free Bi\textsubscript{0.9}Sb\textsubscript{0.1} single layer, and FeCo[17~\AA] on MgO(001), both capped by AlO\textsubscript{x}.

The twin-free Bi\textsubscript{0.9}Sb\textsubscript{0.1} film was deposited at a temperature of -60$^{\circ}$C, with the molecular beam directed along the BaF\textsubscript{2} [$\bar{1}$2${\bar{1}}$] direction, incident at a 40$^{\circ}$ angle from the normal of the substrate, followed by annealing to 200$^{\circ}$C. We attribute the suppression of crystal twinning to the deposition at low temperature and to the tilted molecular beam, which reduce the thermal energy of the atoms on the substrate while imparting them a preferential in-plane direction favoring the growth of a single twinning domain. Deposition at room temperature and normal incidence, on the other hand, produced a Bi\textsubscript{0.9}Sb\textsubscript{0.1} epitaxial film with two twinning domains of similar extension. 
The crystalline quality and suppression of crystal twinning in the epitaxial Bi\textsubscript{0.9}Sb\textsubscript{0.1} film were characterized \textit{in~situ} by reflection high-energy electron diffraction (RHEED) and \textit{ex~situ} by x-ray diffraction (XRD) and atomic force microscopy (AFM), as reported in Figure~\ref{fig:boat1}.

The epitaxy of rhombohedral \mbox{Bi\textsubscript{0.9}Sb\textsubscript{0.1}(0001)} growing on top of cubic \mbox{BaF\textsubscript{2}(111)} is possible due to the lattice matching depicted in Figure~\ref{fig:boat1}(a). We note that Bi\textsubscript{0.9}Sb\textsubscript{0.1} possesses a primitive rhombohedral cell, shown in blue, and  a conventional hexagonal cell, shown in dark gray, to which we refer using the four-axis hexagonal notation (\textit{uvtw}). The in-plane lattice constants are a\textsubscript{BaF\textsubscript{2}}=4.38~\AA~and a\textsubscript{Bi\textsubscript{0.9}Sb\textsubscript{0.1}}=4.54~\AA~ \cite{G278, G257}, leading to a Bi\textsubscript{0.9}Sb\textsubscript{0.1} film with a nominal 3.5$\%$ compressive strain. Owing to the similar size of the Bi and Sb atoms, the Bi\textsubscript{0.9}Sb\textsubscript{0.1} compound is a disordered substitutional alloy. The top of Figure~\ref{fig:boat1}(a) depicts the crystal structure of one twinning domain; its twinned counterpart consists of the same crystal structure rotated by 180$^\circ$ around the rhombohedron diagonal.

The first proof of successful epitaxy is the similarity of the RHEED patterns of the BaF\textsubscript{2}(111) substrate (bottom) and the Bi\textsubscript{0.9}Sb\textsubscript{0.1} film (top) shown in Figure~\ref{fig:boat1}(b); here, the electron beam is maintained along a fixed direction, corresponding to [10$\bar{1}$] for BaF\textsubscript{2} and [2$\bar{1}\bar{1}$0] for Bi\textsubscript{0.9}Sb\textsubscript{0.1}. This result confirms the matching of the in-plane lattice constants, whereas the streaky reflections additionally indicate high crystal quality and surface flatness. The subsequent FeCo layer grows epitaxially in the body-centered cubic phase and twin-free, with the (110) plane oriented out-of-plane, and the [001] direction parallel to the [10$\bar{1}$0] direction of Bi\textsubscript{0.9}Sb\textsubscript{0.1}, as indicated by the corresponding RHEED patterns (see Supporting Information Section~1). The second proof of epitaxy is given by the 2$\theta$/$\omega$ XRD scan shown in Figure~\ref{fig:boat1}(c). The \mbox{twin-free Bi\textsubscript{0.9}Sb\textsubscript{0.1}} bilayer (top in blue) presents reflexes of the BaF\textsubscript{2}(111) and Bi\textsubscript{0.9}Sb\textsubscript{0.1}(0006) planes, which, together with the absence of any undesired reflexes, confirms the film epitaxy. From the full width half maximum of the (0003) reflex, we calculate a crystallite size of 93$\pm$5~\AA for Bi\textsubscript{0.9}Sb\textsubscript{0.1} using the Scherrer equation, which is consistent with the nominal film thickness (see Supporting Information Section~1). The epitaxial FeCo layer is too thin to be detected by XRD.

The suppression of crystal twinning is confirmed by performing azimuthal XRD $\varphi$ scans. Figure~\ref{fig:boat1}(d) shows the unit cells of the BaF\textsubscript{2} substrate (bottom) and Bi\textsubscript{0.9}Sb\textsubscript{0.1} film (top), displaying in color the crystallographic planes probed in the $\varphi$ scans. We investigated the \{002\} planes for BaF\textsubscript{2} and the \{01$\bar{1}$2\} planes for Bi\textsubscript{0.9}Sb\textsubscript{0.1} (shown here are the planes belonging to the twinning domain depicted at the top of Figure~\ref{fig:boat1}(a)). The $\varphi$ scans presented in Figure~\ref{fig:boat1}(e) show that BaF\textsubscript{2} has three-fold rotational symmetry, whereas Bi\textsubscript{0.9}Sb\textsubscript{0.1} has three-fold symmetry shifted by 60$^\circ$ with respect to BaF\textsubscript{2}, as expected from lattice match considerations. On the other hand, a twinned Bi\textsubscript{0.9}Sb\textsubscript{0.1} would have six-fold symmetry, which is not observed. The $\varphi$ scans thus demonstrate the successful deposition of twin-free epitaxial Bi\textsubscript{0.9}Sb\textsubscript{0.1}, growing with the [10$\bar{1}$0] direction parallel to the  [11$\bar{2}$] direction of BaF\textsubscript{2}.

Next, we characterized the topography of an uncapped twin-free Bi\textsubscript{0.9}Sb\textsubscript{0.1} film by AFM (Figure~\ref{fig:boat1}(f)). In the picture, triangular-shaped features can be seen, all similarly oriented, consistently with the absence of twinning. The root-mean-square roughness is 0.37$\pm$0.02~nm. The surface is flatter than previous reports on epitaxial BiSb(0001) with roughness from 0.6 to 1 nm \cite{G197, G254, G255, sasaki2022highly}. This low roughness helps to obtain a high-quality interface with the ferromagnetic layer (FM), as required for studying SOTs accurately. 
Overall, the RHEED, XRD and AFM measurements demonstrate the successful deposition of twin-free epitaxial Bi\textsubscript{0.9}Sb\textsubscript{0.1}(0001) thin films of the highest quality. Previous work reporting twin-free growth of Bi\textsubscript{1-x}Sb\textsubscript{x}(0001) films via MBE was based on a two-step growth method involving high deposition temperatures (150~$^{\circ}$C and 250~$^{\circ}$C) and large film thickness ($\sim$~300~nm) \cite{G257}. Our method involves only a single deposition step at low temperature with a tilted molecular beam and is suited to deposit much thinner films. Other attempts to deposit Bi\textsubscript{1-x}Sb\textsubscript{x}(0001) films using MBE involved Si(111) and GaAs(111) substrates, but, in the first case, no twinning study was conducted \cite{G251}, and, in the second, twinning was reported \cite{G276, G271}. 
}

Finally, the RHEED patterns of the FeCo/Bi\textsubscript{0.9}Sb\textsubscript{0.1} sample deposited on MgO present a cloudy background for both the FeCo and Bi\textsubscript{0.9}Sb\textsubscript{0.1} layers, indicating amorphous growth (see Supporting Information Section~1). The amorphous nature of this film is confirmed by the 2$\theta$/$\omega$ XRD scan in Figure~\ref{fig:boat1}(c) (gray line), where we observe a strong reflex of MgO(001) and small reflexes of non-epitaxial Bi\textsubscript{0.9}Sb\textsubscript{0.1} crystalline phases, indicating the growth of a small fraction of crystalline phases inside a larger amorphous matrix.

\clearpage

\subsection{Magnetoresistance}
 
A first indication of strong charge-spin conversion in Bi\textsubscript{0.9}Sb\textsubscript{0.1} is obtained by performing magnetoresistance measurements. In a single magnetic layer such as FeCo, the longitudinal magnetoresistance varies as $R_{\rm L}=R_{\perp}+\Delta R_{\parallel}m_x^2$, where $R_{\perp}$ is the resistance measured when the unit vector magnetization $\mathbf{m}$ is orthogonal to the current $\mathbf{j} \parallel \mathbf{x}$, and $\Delta R_{\parallel}$ is the increase of resistance when $\mathbf{m}\parallel \mathbf{j}$ \cite{campbell1982chapter}. In nonmagnetic/ferromagnetic bilayers, the spin injection from the nonmagnetic layer into the magnetic layer modifies the conductivity of the entire structure, and the resistance assumes the general form \cite{G41}
\begin{equation}
R_{\rm L}=R_{\rm 0}+\Delta R_{\rm xy}m_x^2+\Delta R_{\rm zy}m_z^2=R_{\rm 0}+\Delta R_{\rm xy}\sin^2\theta\cos^2\varphi+\Delta R_{\rm zy}\cos^2\theta
\label{eq:zero},
\end{equation}
where $R_{\rm 0}$ is the resistance measured when $\mathbf{m}\parallel \mathbf{y}$, $\Delta R_{\rm xy(zy)}$ is the resistance difference between the magnetization pointing along $\mathbf{x}$ ($\mathbf{z}$) and $\mathbf{y}$, and $\theta$ and $\varphi$ are the polar and azimuthal angles of the magnetization, respectively. If the spin current is polarized along $\mathbf{y}$, as is the case for the spin Hall and Rashba-Edelstein effects, the resistance has a minimum for $\mathbf{m} \parallel \pm \mathbf{y}$.

\color{black}

\begin{figure}
  \includegraphics[width=0.9\linewidth]{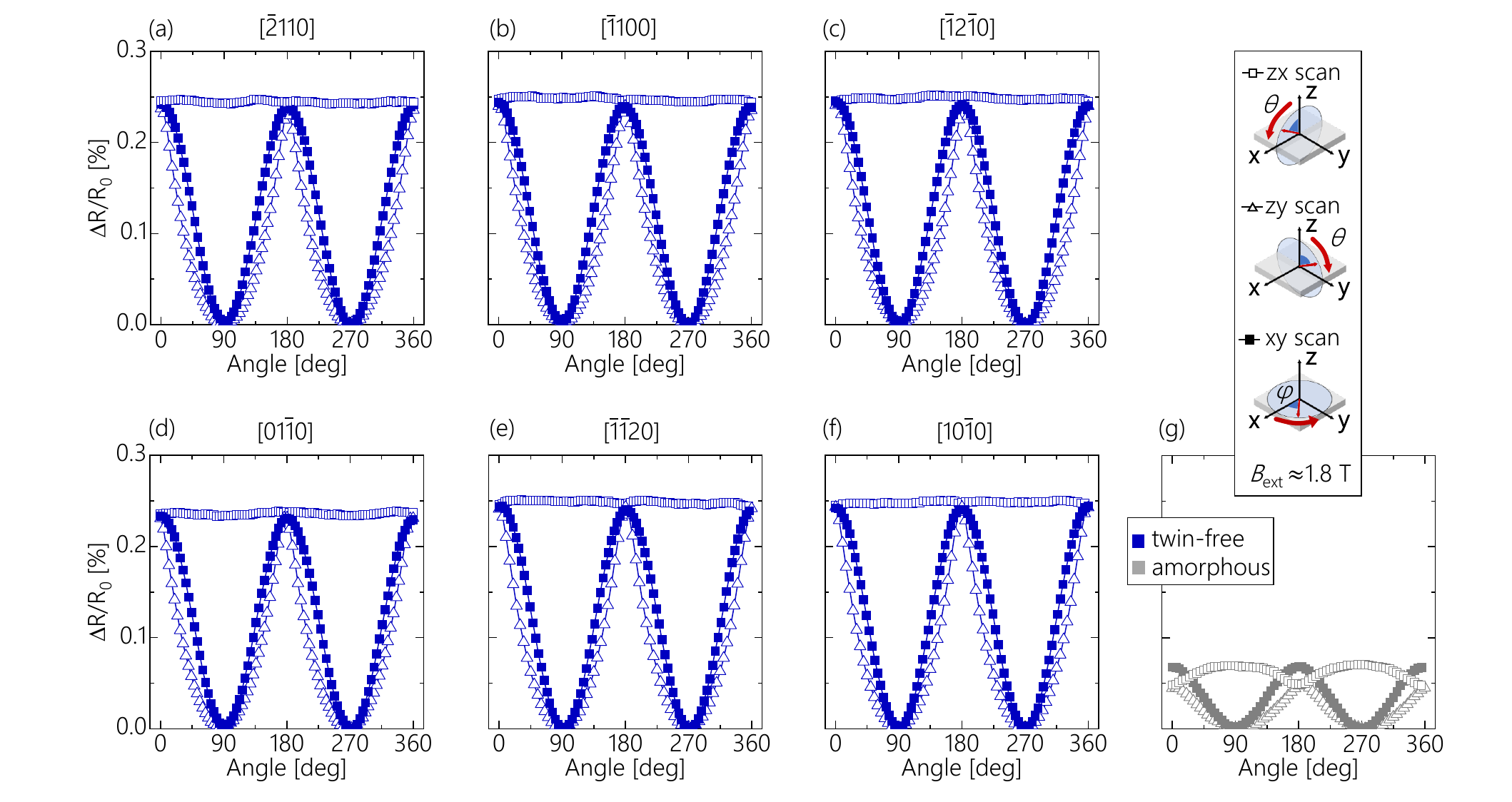}
  \caption{Magnetoresistance of the twin-free Bi\textsubscript{0.9}Sb\textsubscript{0.1}/FeCo bilayer (a-f) and amorphous Bi\textsubscript{0.9}Sb\textsubscript{0.1}/FeCo bilayer (g). The data points represent the percent variation of the longitudinal resistance normalized with respect to $R_{\rm 0}$=$R(\textbf{m}\Vert\textbf{y})$ as the magnetization rotates in the zx (open squares), zy (open triangles), and xy (filled squares) planes. Panels (a-f) show the magnetoresistance when the current $\mathbf{j} \parallel \mathbf{x}$ is injected along different crystallographic directions of Bi\textsubscript{0.9}Sb\textsubscript{0.1}. All measurements were performed at room temperature with current density $j$=1.7$\times$10\textsuperscript{6}~A/cm\textsuperscript{2} by rotating the sample in a constant magnetic field $B$\textsubscript{ext}=1.8~T.
  }
  \label{fig:FIG0000_SMR}
\end{figure}

Figure~\ref{fig:FIG0000_SMR} shows the magnetoresistance of the twin-free and amorphous bilayer samples measured on patterned Hall bars, as described in the Experimental Section. The twin-free Bi\textsubscript{0.9}Sb\textsubscript{0.1}/FeCo bilayer presents a magnetoresistance characterized by $\Delta R_{\rm xy}\approx\Delta R_{\rm zy}>0$ and $\Delta R_{\rm zx}\approx 0$. This behavior is characteristic of the spin Hall magnetoresistance (SMR) in the absence of significant anisotropic magnetoresistance (AMR) \cite{nakayama2013spin, nakayama2016rashba}, and is indicative of strong charge-spin conversion taking place in the bilayer system. More precisely, $\Delta R_{\rm zy}>0$ is the characteristic signature of either the SMR \cite{kim2016spin} or Rashba-Edelstein magnetoresistance \cite{nakayama2016rashba}, whereby the current-induced spin accumulation results in the increase of the bilayer conductivity when the magnetization of the magnetic layer is aligned parallel or antiparallel to the spin polarization of the conduction electrons. As shown later, the charge-spin conversion in our samples is mainly assigned to the bulk carriers of Bi\textsubscript{0.9}Sb\textsubscript{0.1}, and we thus refer to this effect as the SMR. We note also that
$\Delta R_{\rm zx}\approx 0$ and $\Delta R_{\rm xy}\approx\Delta R_{\rm zy}$ are expected when the AMR of the magnetic layer is negligible compared to the SMR. 

The relative magnitude of the SMR, given by $\frac{\Delta R_{\rm xy}}{R_{\rm 0}}\approx 0.25$~\% is as large as in Pt/Co bilayers, the prototypical system for SOT generation  \cite{G117,G41,kawaguchi2018anomalous,G49}, and larger than reported for other TI/FM bilayers. Few observations of SMR-like resistance have been actually reported for such systems, despite their large SOTs. No SMR was found in Bi\textsubscript{2}Se\textsubscript{3}/Y\textsubscript{3}Fe\textsubscript{5}O\textsubscript{12} \cite{lv2022large}
and an SMR of 0.15\% was reported in Bi\textsubscript{2}Se\textsubscript{3}/CoFeB, with a large $\Delta R_{\rm zx}$ contribution that is not present in our epitaxial sample \cite{G25}.
Note that the ordinary magnetoresistance (OMR) can also contribute significantly to the magnetoresistance of TI/FM bilayers, due to the high mobility of TIs.
In 2D systems, the OMR is proportional to the square of the out-of-plane field and appears as \mbox{$\Delta R_{\rm zx}$=$\Delta R_{\rm zy}$ $\neq$ 0}.
The OMR complicates the overall magnetoresistance picture, hindering the estimation of the SMR when the AMR is also present (\mbox{$\Delta R_{\rm xy}$ $\neq$ $\Delta R_{\rm zx}$ $\neq$ 0}).
Magnetoresistance measurements performed as a function of an out-of-plane magnetic field, however, show that the OMR is negligible compared to the SMR in the twin-free Bi\textsubscript{0.9}Sb\textsubscript{0.1}/FeCo bilayer (see Section~3 of the Supporting Information). This allows us to confidently describe this system within the SMR scenario. As the SMR is very sensitive to the quality of the interface and the effective spin Hall angle of the TI, the strong SMR confirms the high quality of our samples and the high conversion efficiency of twin-free Bi\textsubscript{0.9}Sb\textsubscript{0.1}. 

Remarkably, the magnetoresistance of the twin-free sample is nearly independent of the current injection direction, as shown in Figure~\ref{fig:FIG0000_SMR}(a-f) for Hall bar devices oriented along six different crystal directions, one 30$^\circ$ apart from the other. This is a first indication that the charge-to-spin conversion mechanism in crystalline Bi\textsubscript{0.9}Sb\textsubscript{0.1} is isotropic.

In contrast to the twin-free bilayer, the amorphous sample has a significantly smaller magnetoresistance characterized by $\Delta R_{\rm xy}\neq\Delta R_{\rm zy}\neq\Delta R_{\rm zx}\neq 0$, as shown in Figure~\ref{fig:FIG0000_SMR}(g). 
This difference indicates that the charge-spin conversion in amorphous Bi\textsubscript{0.9}Sb\textsubscript{0.1} is strongly reduced compared to the single crystal Bi\textsubscript{0.9}Sb\textsubscript{0.1}, and that there is a contribution from the AMR of FeCo or from the OMR of the amorphous Bi\textsubscript{0.9}Sb\textsubscript{0.1}.
We also find that the amorphous bilayer is more resistive than the twin-free bilayer. The resistivity and magnetization data of both samples are summarized in Section~2 of the Supporting Information.

\color{black}

\clearpage


\begin{figure}
  \includegraphics[width=\linewidth]{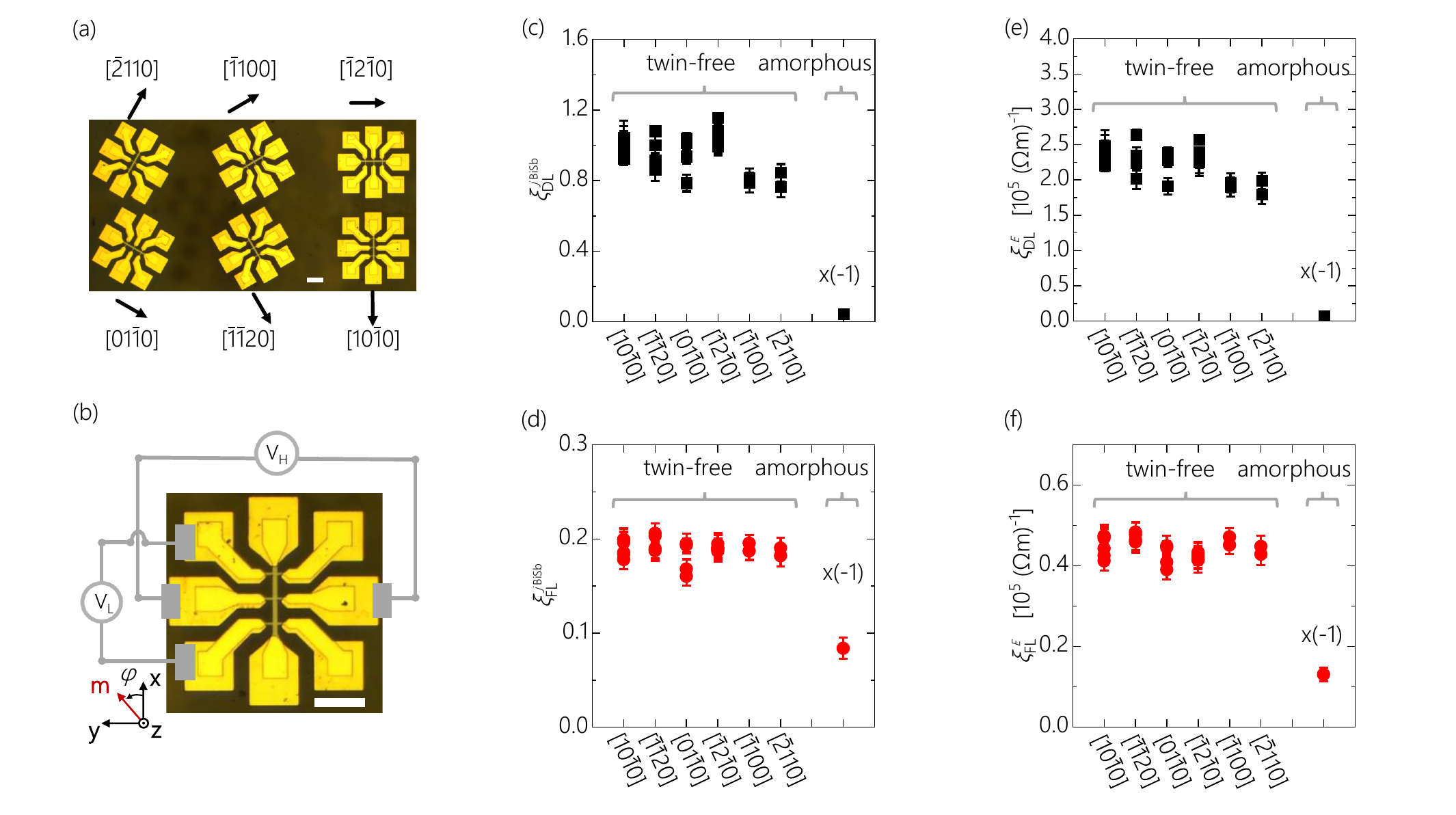}
  \caption{Efficiency of the damping-like and field-like SOTs at room temperature. (a) Optical image of Hall bar devices patterned along different crystallographic directions of the twin-free Bi\textsubscript{0.9}Sb\textsubscript{0.1}/FeCo bilayer. (b) Close up of a device with electrical contacts and coordinate system. The scale bar is 100~$\upmu$m long in both images. \color{black} (c), (d) SOT efficiency with respect to $j_{\rm BiSb}$ for the DL and FL torques in the different samples. \color{black} The direction of the current is indicated for the twin-free Bi\textsubscript{0.9}Sb\textsubscript{0.1}/FeCo bilayer. (e), (f) SOT efficiency with respect to the electric field $E$.
  }
  \label{fig:boat2}
\end{figure}

\clearpage
\color{black} 
\subsection{Spin-orbit torque efficiency}
We quantified the SOTs using the harmonic Hall Voltage method \cite{G50}, accounting for the contribution of the ordinary and anomalous Nernst effects (see Supporting Information Sections~4 and 5). 
We estimated the following effective spin-orbit fields: \mbox{$B\textsubscript{DL}$} for the damping-like torque (DL-SOT), and \mbox{$B\textsubscript{FL+Oe}$}, which includes the Oersted field contribution, for the field-like (FL-SOT). 
To determine ${B_{\rm FL}}$, we subtracted the calculated Oersted field $B_{\rm Oe}$ from ${B_{\rm FL+Oe}}$, taking $B_{\rm Oe}=\frac{\mu_0 {I_{\rm BiSb}}}{2 w}$, where $\mu_0$ is the vacuum permeability, $w$ the current line width and ${I_{\rm BiSb}}$ the current flowing in the nonmagnetic layer. 
${I_{\rm BiSb}}$ was estimated using a parallel resistor model by measuring the longitudinal resistance of the two bilayer samples, a Bi\textsubscript{0.9}Sb\textsubscript{0.1} single film grown on BaF\textsubscript{2}, and a FeCo single film grown on MgO.

In order to compare the SOTs in different systems, it is useful to convert \mbox{$B\textsubscript{DL}$} and \mbox{$B\textsubscript{FL}$} into the so-called SOT efficiencies \cite{G117, G19, G217}.
There are two types of efficiencies: the first normalizes the SOT effective fields to the current density and the second to the applied electric field. The two efficiencies are respectively defined as:
\begin{math}
\xi_{\rm Dl,FL}^{j}=\frac{2e}{\hbar}{M_s}{t_F}\frac{B_{\rm Dl,FL}}{j}
\label{eq:three}
\end{math} 
and 
\begin{math}
\xi_{\rm Dl,FL}^{E}=\frac{2e}{\hbar}{M_s}{t_F}\frac{B_{\rm Dl,FL}}{E},
\label{eq:four}
\end{math}
where $e$ is the elementary charge, $\hbar$ the reduced Planck constant, $j$ the current density, $E=\rho j$ the applied electric field, $\rho$ the resistivity of the bilayer, $t_F$ the thickness of the ferromagnet, and $M_{\rm s}$ the saturation magnetization. By measuring ${B_{\rm DL}}$ and ${B_{\rm FL}}$ separately, we derive the efficiencies $\xi_{\rm DL}$ and $\xi_{\rm FL}$ for the DL- and FL-SOT, respectively. $\xi_{\rm DL}^{j}$ represents the ratio between the amount of spin current absorbed by the FM and the injected charge current, which can be considered as an effective spin Hall angle. $\xi_{\rm DL}^{E}$ represents the ratio between the spin current absorbed by the FM and the applied electric field, that is, the effective spin Hall conductivity. 
For SOTs generated by TIs, it is common to calculate $\xi_{\rm DL,FL}^{j}$ with respect to the current density flowing exclusively through the TI layer, namely $j\textsubscript{BiSb}$. 
Depending on the resistivity of each layer, the values of $j$ and $j\textsubscript{BiSb}$ can differ significantly and therefore affect the efficiencies. For this reason, we report both values of $\xi_{\rm Dl,FL}^{j}$ and $\xi_{\rm Dl,FL}^{j_{\rm BiSb}}$ in Table~\ref{tab:table0_BiSb}.
Accounting for the spin reflection and spin memory loss occurring at the interface between nonmagnetic and magnetic layers, the SOT efficiencies defined above constitute the lower bound for quantifying the charge-to-spin conversion of a nonmagnetic layer.

Figure~\ref{fig:boat2} (a, b) presents images of the Hall bar devices used for the SOT measurements. The room-temperature SOT efficiencies of the twin-free and the amorphous Bi\textsubscript{0.9}Sb\textsubscript{0.1} bilayers are shown in Figure~\ref{fig:boat2}(c-f). For the epitaxial sample, we report the efficiencies as a function of the crystallographic direction parallel to the applied current.
To appreciate the statistical variation of the efficiency, we plot between two and four data points obtained from measuring different devices for each crystallographic direction. 
Regardless of the normalization method, we notice two important results concerning the twin-free Bi\textsubscript{0.9}Sb\textsubscript{0.1} bilayer.
On the one side, there is no evident correlation between the magnitude of the DL- and FL-SOT efficiency and the crystallographic direction along which the current is injected: all the data points appear randomly scattered within a region encompassing $\pm$20\% of the average efficiency.
On the other side, $\xi_{\rm DL}^{j_{\rm BiSb}}$ ($\xi_{\rm DL}^{E}$) reaches almost unity ($2.2\times10\textsuperscript{5}$~$\Omega^{-1}{\rm m}^{^-1}$) and is more than twenty times larger than in the amorphous bilayer. Likewise, $\xi_{\rm FL}^{j_{\rm BiSb}}$ ($\xi_{\rm FL}^{E}$) is significantly larger in the twin-free bilayer compared to the amorphous bilayer. The sign of $\xi_{\rm DL}$ and $\xi_{\rm FL}$ are also inverted between the two samples, which is due to the inverted layer stack.
The average efficiency for all devices and directions is reported in Table~\ref{tab:table0_BiSb}, with uncertainties given by the standard deviation of the data.

\begin{table*}
\centering
\small
\resizebox{\textwidth}{!}{
\begin{tabular}{lllllllllll}
\hline
Sample &$\rho$ &$\frac{\Delta R_{\rm xy}}{R_{0}}$&$\frac{\Delta R_{\rm zy}}{R_{0}}$&$\frac{\Delta R_{\rm zx}}{R_{0}}$ 
&$\xi_{\rm DL}^{\textit{$j$}}$  &$\xi_{\rm FL}^{\textit{$j$}}$
&$\xi_{\rm DL}^{\textit{$j_{\rm BiSb}$}}$  &$\xi_{\rm FL}^{\textit{$j_{\rm BiSb}$}}$ &$\xi_{\rm DL}^{\textit{E}}$ &$\xi_{\rm FL}^{\textit{E}}$  \\
 
 &[$\mu\Omega$cm] &[$\%$]& [$\%$]&[$\%$] 
 &$\times10\textsuperscript{-2}$ &$\times10\textsuperscript{-2}$
 &$\times10\textsuperscript{-2}$ &$\times10\textsuperscript{-2}$ &[10\textsuperscript{4}~($\Omega$m)\textsuperscript{-1}] &[10\textsuperscript{4}~($\Omega$m)\textsuperscript{-1}]\\
\hline

Twin-free  &235 &0.25 &0.25 &$\approx$~0 &52~$\pm$~6 &10~$\pm$~1  &98~$\pm$~11  &19~$\pm$~1     &22~$\pm$~2    &4~$\pm$~0.2\\

Amorphous  &515 &0.07 &0.05 &-0.02       &-3~$\pm$~1  &-6~$\pm$~1 &-4~$\pm$~1   &-7.5~$\pm$~1   &-0.6~$\pm$~0.2    &-1~$\pm$~0.2\\

\hline
\end{tabular}
}

\caption{\label{tab:table0_BiSb} Resistivity, magnetoresistance, and SOT efficiency of the twin-free and amorphous Bi\textsubscript{0.9}Sb\textsubscript{0.1} samples at room temperature. The normalized magnetoresistance is reported for xy, zy and zx angular scans, the DL- and FL-SOT efficiency are normalized to the total current density ($j$), to the current density through the Bi\textsubscript{0.9}Sb\textsubscript{0.1} layer ($j_{\rm BiSb}$), and to the electric field ($E$). The uncertainty represents the standard deviation of the data measured on different devices.
}
\end{table*}

We now compare the twin-free and amorphous bilayers by correlating their SOT and magnetotransport properties and discuss them in relation with other SOT-generating materials and similar Bi\textsubscript{1-x}Sb\textsubscript{x} compounds reported in the literature. The twin-free epitaxial bilayer sample presents SOT efficiencies significantly larger than the amorphous bilayer. A similar trend occurs for the SMR (${\Delta R_{\rm zy}}$). 
From these observations, we conclude that the crystalline quality is crucial for achieving a large charge-to-spin conversion in Bi\textsubscript{0.9}Sb\textsubscript{0.1}. \color{black} Additionally, the ferromagnetic interface of the amorphous Bi\textsubscript{0.9}Sb\textsubscript{0.1} bilayer is rougher than that of the twin-free epitaxial bilayer, as RHEED images indicate. The interface quality, which is affected also by the substrate and stacking order, is another factor that can influence the charge-to-spin conversion and the interface transparency to the spin current.
\color{black}

The SOT efficiency of the twin-free Bi\textsubscript{0.9}Sb\textsubscript{0.1} bilayer compares favorably with the 5$d$ heavy metals. Pt, the most investigated material for SOT generation, has reported efficiencies per total current density of \mbox{$\xi_{\rm DL,Pt}^{j}=0.06-0.2$} and \mbox{$\xi_{\rm FL,Pt}^{j}=-0.07$}, and effective spin Hall conductivities of $\xi_{\rm DL,Pt}^{E}=2-4\times10^{5}$~($\Omega$m)\textsuperscript{-1}, $\xi_{\rm FL,Pt}^{E}=-2\times10^{5}$~($\Omega$m)\textsuperscript{-1} \cite{G117,G19}.
For the same stacking order of the nonmagnetic and magnetic layers, the DL-SOT has the same sign as in Pt, whereas the FL-SOT has opposite sign \cite{G50}. 
Due to the larger resistivity of Bi\textsubscript{0.9}Sb\textsubscript{0.1} compared to Pt, the SOT efficiencies normalized to the electric field are smaller than Pt. Compared to other TIs, however, Bi\textsubscript{0.9}Sb\textsubscript{0.1} has a larger electrical conductivity, representing one of the best materials that combines a large SOT efficiency with relatively low resistivity \cite{G126, G256, G253}. SOT studies using the harmonic Hall voltage method reported efficiencies $\xi_{\rm DL,FL}^{j\textsubscript{TI}}$ spreading from 0.15 to 18 for Bi-based chalcogenides and other TIs \cite{G211, binda2021spin,G123,G13, G130, G132, G129, wu2019spin}. The SOT efficiency measured in these systems varies significantly depending on the TI composition, current density normalization, and measurement method.

Table~\ref{tab:table0_BiSbliterature} compares the SOT efficiency of heterostructures based on Bi\textsubscript{1-x}Sb\textsubscript{x} reported in the literature. We focus here on the DL-SOT, which received more attention due to its primary role in inducing magnetization switching and domain wall motion \cite{G117}. Also in this case, we observe a significant spread of $\xi_{\rm DL}^{j}$ and $\xi_{\rm DL}^{E}$, which can be explained by different considerations. The largest SOT efficiency of $\xi_{\rm DL}^{j\textsubscript{BiSb}}=52$ has been reported for epitaxial Bi\textsubscript{0.9}Sb\textsubscript{0.1}($\bar{1}$2$\bar{1}$6) deposited by MBE on GaAs(001)/Mn\textsubscript{0.6}Ga\textsubscript{0.4} \cite{G126}. This very large SOT efficiency has been measured indirectly using the current-induced shift of hysteresis loops and attributed to the SSs of the ($\bar{1}$2$\bar{1}$6) plane \cite{zhu2013three}. Measurements of textured  Bi\textsubscript{0.85}Sb\textsubscript{0.15}(11$\bar{2}$0)/[Co/Pt] multilayers deposited by sputtering also gave a remarkable $\xi_{\rm DL}^{j\textsubscript{BiSb}}=10.7$ \cite{G256}. Studies of sputtered  Bi\textsubscript{1-x}Sb\textsubscript{x} films report $\xi_{\rm DL}^{j\textsubscript{BiSb}}$ of order unity \cite{G126, G197, G270, G254, huy2023large}, whereas measurements of epitaxial Bi\textsubscript{0.74}Sb\textsubscript{0.26}(0001) with a ferromagnetic layer grown ex situ \cite{G261} gave a negligible efficiency and spin pumping measurements of FeGaB/BiSb gave an efficiency of 1\% \cite{sharma2021light}. 
The large difference of $\xi_{\rm DL}^{j\textsubscript{BiSb}}$ reported in the literature were initially attributed to the difference in the crystalline quality and orientation of the samples. Our comparative investigation of samples grown in the same environment and measured using the same technique shows that single-crystal Bi\textsubscript{0.9}Sb\textsubscript{0.1}(0001) films have a much larger $\xi_{\rm DL}^{j\textsubscript{BiSb}}$ compared to amorphous Bi\textsubscript{0.9}Sb\textsubscript{0.1} and confirm the importance of high crystalline quality to reach large conversion efficiency. However, the DL-SOT efficiency is not as large as reported previously for Bi\textsubscript{0.9}Sb\textsubscript{0.1}($\bar{1}$2$\bar{1}$6) films.

\color{black}
\begin{table*}
\centering
\small
\resizebox{\textwidth}{!}{
\begin{tabular}{lllllll}
\hline

Composition &Heterostructure &Deposition technique &Crystal structure &Anisotropy, &$\xi_{\rm DL}^{\textit{$j_{\rm BiSb}$}}$ &$\xi_{\rm DL}^{\textit{E}}$ \\
&[Layer thickness, nm] &  &  &Method  &  &[10\textsuperscript{5}~($\Omega$m)\textsuperscript{-1}]\\
\hline

Bi\textsubscript{0.9}Sb\textsubscript{0.1} [this work]
&BaF\textsubscript{2}(111)/Bi\textsubscript{0.9}Sb\textsubscript{0.1}[10]/FeCo[1.7]
 &MBE
 & Epi (0001), twin-free &IPMA, HHV &1 &2.2  \\

Bi\textsubscript{0.9}Sb\textsubscript{0.1} \cite{G126} &GaAs(001)/Mn\textsubscript{0.6}Ga\textsubscript{0.4}[3]/Bi\textsubscript{0.9}Sb\textsubscript{0.1}[10] &MBE &
Epi ($\bar{1}$2$\bar{1}$6)
&PMA, HLS &52 &130 \\

Bi\textsubscript{0.74}Sb\textsubscript{0.26} \cite{G261, G251} 
&Si(111)/Bi\textsubscript{0.74}Sb\textsubscript{0.26}[4]/Co[4] 
& MBE, air transfer; evaporation &Epi (0001) 
&IPMA, HHV 
&$\approx$~0 &$\approx$~0 \\

Bi\textsubscript{0.83}Bi\textsubscript{0.17} \cite{G197}
 &SiO\textsubscript{2}/Ta[0.5]/Te[2]/Bi\textsubscript{0.83}Sb\textsubscript{0.17}[10]/CoFeB[2]
 &Sputtering of Bi/Sb multilayer &Epi (0001) 
 &IPMA, HHV 
 &1.2 &0.6 \\

Bi\textsubscript{0.85}Sb\textsubscript{0.15} \cite{G259}
&
SiO\textsubscript{2}/MgO[10]/Pt[0.8]/Co[0.6]/Pt[0.8]/Bi\textsubscript{0.85}Sb\textsubscript{0.15}[10]
 &Sputtering
 &Unknown &PMA, HHV &2.4   &  \\

Bi\textsubscript{0.85}Sb\textsubscript{0.15} \cite{G270}
 &SiO\textsubscript{2}/CoTb[2.7]/Pt[1]/Bi\textsubscript{0.85}Sb\textsubscript{0.15}[10]
 &Sputtering, air transfer, MBE
 &Textured (0001)
 &PMA, HHV
 &3.2 &  \\

Bi\textsubscript{0.85}Sb\textsubscript{0.15} \cite{G256}
&Al\textsubscript{2}O\textsubscript{3}(0001)/(Co[0.4]/Pt[0.4])\textsubscript{2}/Bi\textsubscript{0.85}Sb\textsubscript{0.15}[10]
 &Sputtering
 &Textured (11$\bar{2}$0) 
 &PMA, HHV  &10.7 &16  \\

\color{black}
Bi\textsubscript{0.85}Sb\textsubscript{0.15} \cite{khang2022nanosecond}
&\color{black}Si/SiO\textsubscript{x}/Ins. buffer/(Co[0.4]/Pt[0.4])\textsubscript{2}/Bi\textsubscript{0.85}Sb\textsubscript{0.15}[10]
 &\color{black}Sputtering
 &\color{black}Unknown 
 &\color{black}PMA, HHV  &3.5 &  \\
\color{black}

Bi\textsubscript{0.85}Sb\textsubscript{0.15} \cite{G254}
&Al\textsubscript{2}O\textsubscript{3}(0001)/Bi\textsubscript{0.85}Sb\textsubscript{0.15}[10]/Ru[1]/Co[1]/ Pt[1] &Sputtering
 &Unknown &PMA, HHV &1.7 &  \\ 

Bi\textsubscript{0.85}Sb\textsubscript{0.15} \cite{sharma2021light}
&Al\textsubscript{2}O\textsubscript{3}(0001)/FeGaB[6]/Bi\textsubscript{0.85}Sb\textsubscript{0.15}[10] &Sputtering
 &Polycrystalline &IPMA, SP &0.01 & \\

\hline

\end{tabular}
}
\caption{\label{tab:table0_BiSbliterature} Summary of the different Bi\textsubscript{1-x}Sb\textsubscript{x} compounds investigated for SOT generation. Composition, deposition technique, crystal structure, magnetic anisotropy (perpendicular: PMA, in-plane: IPMA), measurement method (HHV: harmonic Hall voltage, HLS: hysteresis loop shift, SP: spin pumping), $\xi_{\rm DL}^{j\textsubscript{BiSb}}$ and $\xi_{\rm DL}^{E}$ at room temperature.
}
\end{table*}

\begin{figure}
  \includegraphics[width=0.5\linewidth]{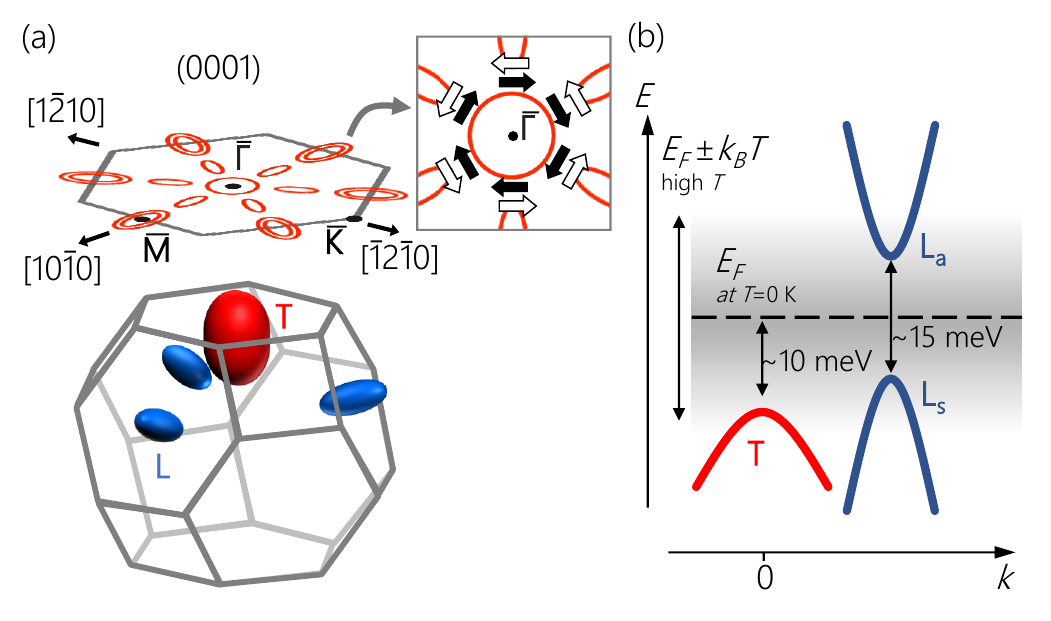}
  \caption{
(a) 3D Brillouin zone of Bi\textsubscript{0.9}Sb\textsubscript{0.1} with bulk conduction pockets at the Fermi surface (bottom), and projected 2D Brillouin zone on the (0001) plane showing the Fermi contours of the SSs (top). The arrows in the inset depicts the spin polarization of the SSs near the $\bar{\Gamma}$ point. (b) Schematic band structure of Bi\textsubscript{0.9}Sb\textsubscript{0.1} showing the dispersion of the bulk bands nearby the T and L points. The energy gaps correspond to those of bulk Bi\textsubscript{0.9}Sb\textsubscript{0.1}\cite{G283}, which represent a lower limit considering the gap opening effect due to quantum confinement in thin films.  
  }
  \label{fig:boat0}
\end{figure}

In order to shed light on the mechanisms at the origin of the SOTs, we discuss the SOTs of the twin-free Bi\textsubscript{0.9}Sb\textsubscript{0.1} bilayer in relationship with the surface and bulk band structure of Bi\textsubscript{0.9}Sb\textsubscript{0.1} obtained from ab-initio electronic structure calculations and photoemission measurements \cite{G265, G274}.
The bulk Fermi surface of Bi\textsubscript{0.9}Sb\textsubscript{0.1} presents a hole pocket at the T point of the Brillouin zone and hole and electron pockets at the L points, as sketched in Figure~\ref{fig:boat0}(a) \cite{G52,  G283,G274, G249}. 
The valence and conduction bands at the L points are separated by a small energy gap of 15~meV \cite{G283}; furthermore, the band symmetry is inverted with respect to a conventional insulator, making it topological.
Figure~\ref{fig:boat0}(b) shows the schematic dispersion of the bulk band states near the T and L points, with the Fermi level positioned in the middle of the L-point energy gap, as for an ideal semiconductor. 
The small gap allows for the excitation of bulk carriers at high enough temperature.
Additionally, Bi\textsubscript{0.9}Sb\textsubscript{0.1} possesses highly anisotropic SSs on the (0001) plane \cite{G201, G265, G275}, sketched on the projected Brillouin zone in Figure~\ref{fig:boat0}(a) \cite{G265}.
Spin-polarized ARPES has shown that the isotropic SSs enclosing the $\bar{\rm{\Gamma}}$ point present a clockwise in-plane spin polarization. 
In contrast, the anisotropic SSs along the $\bar{\rm{\Gamma}}$-$\bar{\rm{M}}$ line present a spin polarization that points counterclockwise around the $\bar{\rm{\Gamma}}$ point with a six-fold rotational symmetry (see inset in Figure~\ref{fig:boat0}) \cite{G201, G265, G275}.

At room temperature, SOTs can be generated by both the spin-polarized SSs and the thermally excited bulk carriers.
Our result shows that the SOTs are isotropic at room temperature, , independent of the direction of the injected current in the (0001) plane. The same isotropic behavior appears in the  SMR. This evidence indicates that charge-spin conversion originates from the isotropic SSs or bulk carriers with isotropic spin-momentum locking near the $\bar{\Gamma}$ or T point.

\color{black}

\begin{figure}
  \includegraphics[width=\linewidth]{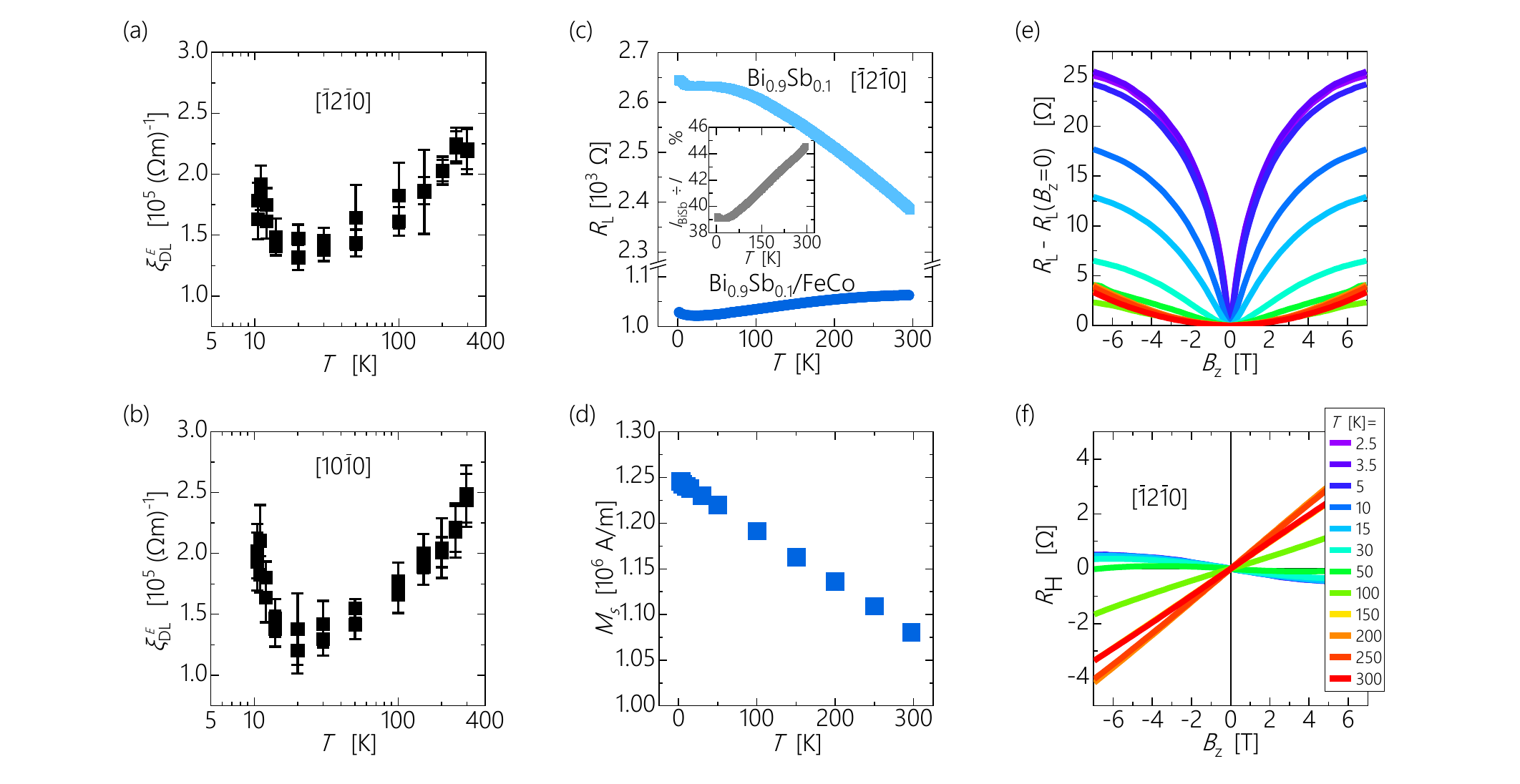}
  \caption{SOT efficiency and magnetotransport properties of the twin-free Bi\textsubscript{0.9}Sb\textsubscript{0.1} bilayer and single layer samples as a function of temperature. (a), (b) SOT efficiency with respect to $E$ as a function of temperature for the device oriented along [$\bar{1}$2$\bar{1}$0] and [10$\bar{1}$0], respectively. To account for Joule heating, the temperature was calibrated by comparing the longitudinal resistances measured using a small current density, $j$=4.2$\times$10\textsuperscript{4}~A/cm\textsuperscript{2} and the SOT measurement current, $j$=1.7$\times$10\textsuperscript{6}~A/cm\textsuperscript{2} (see Supporting Information Section~6). (c) Longitudinal resistance versus temperature of the twin-free Bi\textsubscript{0.9}Sb\textsubscript{0.1} bilayer and the single layer samples at \mbox{$j\approx$4$\times$10\textsuperscript{4}~A/cm\textsuperscript{2}} for devices oriented along [$\bar{1}$2$\bar{1}$0]. (d) Saturation magnetization as a function of temperature. (e), (f) Longitudinal and Hall resistance of a twin-free Bi\textsubscript{0.9}Sb\textsubscript{0.1} single layer with current applied along  [$\bar{1}$2$\bar{1}$0]. 
 }
  \label{fig:boat4}
\end{figure}

\clearpage

\subsection{Spin-orbit torque efficiency versus temperature}

To gain further insight into the origin of the SOTs, we studied the DL-SOT efficiency as a function of temperature. The conductivity of the bulk carriers in a TI with a small bandgap such as Bi\textsubscript{0.9}Sb\textsubscript{0.1} is expected to diminish with decreasing temperature. Thus, the topological SSs should increasingly dominate the transport at low temperature, which allows for separating the bulk and SS contributions. Figure~\ref{fig:boat4}(a-d) shows the temperature dependence of $\xi_{\rm DL}^{E}$ measured along two non-equivalent high-symmetry directions together with the longitudinal resistance $R_{\rm L}$ and saturation magnetization $M\textsubscript{s}$ of the twin-free Bi\textsubscript{0.9}Sb\textsubscript{0.1} bilayer. Additionally, we also characterized $R_{\rm L}$ and the Hall resistance $R_{\rm H}$ of a twin-free Bi\textsubscript{0.9}Sb\textsubscript{0.1} single layer.

The SOT efficiency accounts for the \color{black} temperature-dependent $M\textsubscript{s}$ (Figure~\ref{fig:boat4}(d)) \color{black}, thus reflecting changes in the charge-spin conversion efficiency only. Within the error of our measurements, we observe a similar behavior for devices oriented along the [$\bar{1}$2$\bar{1}$0] and [10$\bar{1}$0] directions: $\xi_{\rm DL}^{E}$ decreases with decreasing temperature by about 40-50\% until a minimum at about 20~K, followed by a sharp increase at a lower temperature, which rises almost up to the room-temperature value. The SOT generation is isotropic at low temperatures as it is at room temperature. 

As a first step to analyze the temperature dependence of the SOT efficiency, we measured the temperature dependence of $R_{\rm L}$ for both the Bi\textsubscript{0.9}Sb\textsubscript{0.1} bilayer and the single layer samples, shown in Figure~\ref{fig:boat4}(c).
Because the resistance of devices oriented along different crystallographic directions has a similar trend, we present only the data relative to the [$\bar{1}$2$\bar{1}$0] direction. In the bilayer, $R_{\rm L}$ decreases by up to 5$\%$ until about 25~K, followed by a slight increase. The decrease of resistance with temperature is attributed to the resistivity of the metallic CoFe layer. The increase at low temperature could either be due to the Kondo effect in FeCo \cite{he2005resistivity} or to the increase of resistivity of the Bi\textsubscript{0.9}Sb\textsubscript{0.1} layer. We estimated the percentage of current flowing through Bi\textsubscript{0.9}Sb\textsubscript{0.1} in the bilayer sample using a parallel resistor model. The inset of Figure~\ref{fig:boat4}(c) shows that the current in the Bi\textsubscript{0.9}Sb\textsubscript{0.1} layer decreases by no more than 5$\%$ from 300 to 2~K, which cannot explain the SOT variation as a function of temperature observed in our samples.
The single layer sample has a semiconductor-like behavior with a monotonous resistance increase of 15$\%$ down to 50~K, a plateau between 50 and 15~K and a steep increase below 10~K. The resistance plateau is usually associated with a dominant conduction contribution of the SSs \cite{G154}. A very similar semiconductor-to-metal transition was also observed in Bi(0001) thin films and associated with the presence of SSs \cite{xiao2012bi, kroger2018controlling}. The upturn below 10~K is attributed to the quantum correction of the electron-electron interaction \cite{kroger2018controlling}. The conductivity of our sample at 270~K (2.1$\times$10$^5~(\Omega$m)$^{-1}$) is close to that reported for films of similar thickness and stoichiometry (1-4$\times$10$^5~(\Omega$m)$^{-1}$) \cite{G276}.   

 Figure~\ref{fig:boat4}(e) presents $R_{\omega, \rm L}$ and (f) $R_{\rm H}$ at various temperatures for a single layer device oriented along [$\bar{1}$2$\bar{1}$0] and current density $j$=5$\times$10\textsuperscript{4}~A/cm\textsuperscript{2}. At room temperature, we observe an ordinary magnetoresistance effect with $R_{\rm L} \propto B^{2}_{\rm z}$, whereas below about 30~K the weak antilocalization effect becomes dominant. The weak antilocalization curves were fitted using the Hikami-Larkin-Nagaoka equation \cite{hikami1980spin} (see Supporting Information Section~7). Our analysis indicates that there is only one channel that contributes to the weak antilocalization at low temperature, whereas in a TI like Bi\textsubscript{0.9}Sb\textsubscript{0.1} we should expect multiple channels corresponding to all the topological SSs. This apparent discrepancy can be explained by the coupling of the topological SSs mediated by the conducting bulk states, leading to only one coherent transport channel. This bulk-surface coupling has been reported in TIs \cite{G227} as well as in epitaxial Bi films \cite{brahlek2015transport}.

With decreasing temperature, the slope of $R_{\rm H}$ changes from positive to negative, indicating a change of the majority carrier type from bulk hole carriers to electron carriers, either from the bulk or the surface. The change of the majority carriers is also observed in the bilayer at temperatures higher than 150~K, suggesting that the transport properties of Bi\textsubscript{0.9}Sb\textsubscript{0.1} are not strongly affected by the deposition of FeCo. From the increase of $R_{\rm L}$ and the change of slope of $R_{\rm H}$ we conclude that charge transport is dominated by holes at high temperature and that both electrons and holes contribute at low temperature, with the total number of hole carriers decreasing at low temperature. These conclusions are in good agreement with the behavior expected from a small band gap semiconductor like Bi\textsubscript{0.9}Sb\textsubscript{0.1} \cite{G287} or from a semimetal like Bi \cite{G285, G286, G288, G289}.

The picture that emerges from our measurements is that the SOT above 20~K are associated with thermally-excited bulk hole carriers at the T point of the Brillouin zone, in agreement with the SMR and SOT measurements evidencing isotropic charge-spin conversion. Our results are also in line with calculations of the spin Hall conductivity in BiSb \cite{G284}, where the T point contribution gives positive spin Hall conductivity, whereas the contribution of the L point is negative. Consequently, the lower the temperature, the smaller the T point contribution to the transport, resulting in lower SOT efficiency.
A previous study of the temperature dependence of SOTs in sputtered Bi\textsubscript{1-x}Sb\textsubscript{x} films also reported a correlation between the amount of thermal carriers and SOT efficiency, but, differently from our work, attributed the thermal carriers to electrons near the L point \cite{G197}. We tentatively explain this discrepancy with the unintentional doping in the sputtered film caused by crystal defects, which brings the Fermi level near or into the conduction band at all temperatures.

Furthermore, the DL-SOT efficiency of the twin-free bilayer increase again below 20~K along both inequivalent crystallographic axis of Bi\textsubscript{0.9}Sb\textsubscript{0.1}. This strong increase corresponds neither to the partial suppression of the bulk contribution to the conductivity observed at around 50~K nor to the change of the main carrier contribution from hole to electron occurring at the same temperature. Instead, the SOT increase occurs when weak antilocalization starts to dominate the magnetoresistance concurrently with the increase of the phase coherence length (see Supporting Information Section~7). From the temperature dependence of the phase coherence length we conclude that the source of dephasing is electron-electron scattering, which decreases at low temperature. While the specific role of the scattering on the SOT in topological insulator heterostructures is not fully understood, inelastic electron-electron scattering is expected to be detrimental to the spin polarization of the SSs and the resulting torques \cite{G290}. The increase of the spin polarization of the SSs at low temperature has been previously proposed to explain a similar increase of the SOT below 10~K in magnetic topological insulator/ topological insulator bilayer \cite{G222}. The weak antilocalization effect would also reduce the resistivity of Bi\textsubscript{0.9}Sb\textsubscript{0.1} leading to more efficient generation of SOT. However, this effect corresponds only to a 1-2\% resistance change, which can hardly explain the sharp increase of the SOT. We conclude that the upturn of the SOT efficiency below 20~K is due to the increased polarization of the isotropic SSs around the $\bar{\Gamma}$ point at low temperature.

\section{Conclusion}

In summary, we have grown high-quality twin-free epitaxial layers of Bi\textsubscript{0.9}Sb\textsubscript{0.1}(0001) using a single-step oblique MBE deposition method at low temperature and investigated the magnetoresistance and SOTs of Bi\textsubscript{0.9}Sb\textsubscript{0.1}/FeCo bilayers as a function of current direction and temperature.
The epitaxial Bi\textsubscript{0.9}Sb\textsubscript{0.1}(0001) films present a semiconducting behavior, with charge transport dominated by holes above 50~K and both electrons and holes at low temperature.
The magnetoresistance of epitaxial twin-free Bi\textsubscript{0.9}Sb\textsubscript{0.1}/FeCo bilayers shows a sizable SMR-like behavior with $\frac{\Delta R_{\rm xy}}{R_{\rm 0}}\approx 0.25$~\%, indicative of strong charge-spin conversion, unlike that of amorphous bilayers. The charge-to-spin conversion efficiency derived from measurements of the DL-SOT is about 1 in the epitaxial Bi\textsubscript{0.9}Sb\textsubscript{0.1} and less than 0.1 in amorphous Bi\textsubscript{0.9}Sb\textsubscript{0.1}, showing that the crystalline quality is key to obtain strong spin injection from this material. Both the SMR and SOTs are isotropic in the epitaxial twin-free Bi\textsubscript{0.9}Sb\textsubscript{0.1} bilayers, despite the highly anisotropic SSs existing on the (0001) plane. Additionally, the SOT efficiency decreases strongly from 300~K to 20~K. 
These findings indicate that, down to 20~K, the SOTs originate mainly from thermally-excited hole carriers around the isotropic T-point of the Brillouin zone in the bulk of the material.
Finally, below 20~K the DL-SOT efficiency increases, reaching close to the room-temperature value at 10.5~K. Based on weak antilocalization measurements, we attribute this upturn to the increased spin polarization of the isotropic SSs around the $\bar{\Gamma}$ point of the Brillouin zone due to reduced electron scattering.
Our work demonstrates a novel route to grow untwinned topological insulators and sheds light on the magnitude of charge-spin conversion and the mechanisms at the origin of SOTs in epitaxial Bi\textsubscript{0.9}Sb.

\clearpage

\section{Experimental Section}

\threesubsection{Material Growth}\\
The films were deposited by MBE in an ultra-high-vacuum system with base pressure $<$10\textsuperscript{-10}~mbar. 
High purity elements ($\geq$ 99.999$\%$) were evaporated from Knudsen cells with the following growth rates monitored by a quartz crystal balance: Bi (3.24~\AA/min), Sb (0.31~\AA/min), Fe (0.23~\AA/min), Co (0.21~\AA/min), Al (0.9~\AA/min). 
The elements Bi, Sb, and Fe, Co were co-deposited. A capping Al layer was deposited with an O\textsubscript{2} pressure of 1.6$\times$10\textsuperscript{-6}~mbar on all samples to protect them from air exposure. 
The MgO(001) and BaF\textsubscript{2}(111) substrates were sonicated in acetone and isopropanol for 5~minutes before being inserted
into the vacuum chamber and heated to about 600$^\circ$C for 90~minutes. 
The twin-free Bi\textsubscript{0.9}Sb\textsubscript{0.1} film was deposited at a temperature of -60$^{\circ}$C, with the molecular beam directed along the BaF\textsubscript{2} [$\bar{1}$2${\bar{1}}$] direction, at a 40$^{\circ}$ angle from the surface normal. 
We attribute the suppression of crystal twinning to the reduced random thermal energy of the adatoms on the substrate held at low temperature and to the anisotropic in-plane velocity induced by the oblique deposition, which favor the growth of a single twinning crystal domain. 
MBE at normal incidence at room temperature produced a Bi\textsubscript{0.9}Sb\textsubscript{0.1} epitaxial film possessing two twinning crystal domains with similar extension (not shown). 
After deposition, the Bi\textsubscript{0.9}Sb\textsubscript{0.1} film was annealed at about 200$^\circ$C for 30 minutes. 
All the other layers were deposited at room temperature, with the molecular beam at normal incidence, including the amorphous Bi\textsubscript{0.9}Sb\textsubscript{0.1} layer.

\threesubsection{Device Fabrication}\\
We fabricated Hall bar devices of 10~$\mu$m width and 100~$\mu$m length. We performed a first step of UV-photolithography to deposit photoresist patches of the same shape as the devices, followed by an  Ar-milling step to etch the uncovered area nearby the devices down to the substrate. We performed a second step of UV-photolithography, followed by a thermal evaporation step, in order to deposit metallic contact pads of Ti~[10~nm]/Au~[200~nm] by lift-off.

\threesubsection{Magnetotransport measurements}\\

We characterized the anomalous Hall resistance ($R$\textsubscript{AHE}) and the saturation field $B$\textsubscript{sat} of the bilayer samples by using an alternate current and measuring the 1\textsuperscript{st} harmonic Hall signal ($R_{\rm H}$)) as a function of an out-of-plane magnetic field (see Supporting Information).  We obtained similar results for the differently-oriented devices of the twin-free Bi\textsubscript{0.9}Sb\textsubscript{0.1} bilayer sample. We also characterized the magnetoresistance by measuring the 1\textsuperscript{st} harmonic longitudinal resistance ($R_{\rm L}$) by rotating the devices in the xy, zy, zx planes in a static magnetic field of 1.8~T, which is large enough to saturate the magnetization along any direction in all the samples, as shown in Figure~\ref{fig:FIG0000_SMR}. 

For the magnetotransport and SOT measurements of the twin-free Bi\textsubscript{0.9}Sb\textsubscript{0.1} bilayer and single layer samples we used Hall bars oriented along six different Bi\textsubscript{0.9}Sb\textsubscript{0.1} crystal directions, one 30$^\circ$ apart from the other, as shown in Figure~\ref{fig:boat2}(a).
At room temperature, the devices were wire-bonded and mounted on a motorized stage allowing for in-plane and out-of-plane rotations in an electromagnet producing fields up to 1.8~T. 
Using the reference system shown in Figure~\ref{fig:boat2}(b), we define $\varphi$ as the azimuthal angle between the x-axis and the magnetization vector $\textbf{m}$, and $\theta$ as the polar angle between the z-axis and $\textbf{m}$. 
The measurements were performed using an a.c. current and recording the 1\textsuperscript{st} and 2\textsuperscript{nd} harmonic of the transverse (Hall) and longitudinal resistances as a function of the magnetic field direction (angle scan) or amplitude (field sweep). 
We used current densities in the range of $j$=(0.7-1.7)$\times$10\textsuperscript{6}~A/cm\textsuperscript{2} with a frequency of $\omega$/2$\pi$=10~Hz. The current density $j$ is calculated by considering the total thickness of the entire stack, excluding the AlO\textsubscript{x} capping. We measured the resistivities of the bilayer and the single layer samples, and then, via a parallel resistor model, we calculated the resistivities of the other layers constituting the heterostructures. 
The data are shown in Section~1 of the Supporting Information.

\clearpage

\medskip
\textbf{Supporting Information} \par 
Supporting Information is available from the Wiley Online Library or from the author.

 \medskip
 \textbf{Acknowledgements} \par 
The authors thank Charles-Henri Lambert for performing the measurements with the superconducting quantum interference device (SQUID).
This research was supported by the Swiss National Science Foundation (Grant No. 200020-200465). P.N. acknowledges the support of the ETH Zurich Postdoctoral Fellowship Program (Grant No. 19-2 FEL-61).

\medskip

%

\bibliographystyle{MSP}
\bibliography{MSP-template}



\providecommand{\noopsort}[1]{}\providecommand{\singleletter}[1]{#1}%

\end{document}